\documentclass[useAMS,usenatbib,usegraphicx]{mn2e}

\usepackage{aas_macros}
\usepackage{amssymb,amsmath}

\title[MCAO observations of Trumpler\,14]{A benchmark for multi-conjugated AO: VLT--MAD observations of the young massive cluster Trumpler\,14 \thanks{Based on Observations with VLT--MAD collected during the Science Verification campaign at the European Southern Observatory, Paranal, Chile.}}
\author[Rochau et al.]{B. Rochau$^{1}$\thanks{E-mail:
rochau@mpia.de}, W. Brandner$^{1}$, A. Stolte$^{2}$, T. Henning$^{1}$, N. Da Rio$^{1,3}$,\\\\{\rm \LARGE M. Gennaro$^{1}$, F. Hormuth$^{1}$, E. Marchetti$^{4}$ and P. Amico$^{4}$}\\
$^{1}${Max-Planck-Institut f{\"u}r Astronomy, K{\"o}nigstuhl 17, 69115 Heidelberg}\\
$^{2}${Argelander Institut f{\"u}r Astronomie, Universit{\"a}t Bonn, Auf dem H{\"u}gel 71, 53121 Bonn}\\
$^{3}${Space Telescope Science Institute, 3700 San Martin Drive, Baltimore, MD 21218}\\
$^{4}${European Southern Observatory, Karl-Schwarzschild-Stra\ss e 2, 85748 Garching bei M{\"u}nchen}}
\begin{document}

\date{Accepted 2011 July 30.  Received 2011 July 30; in original form 2011 May 8}

\pagerange{\pageref{firstpage}--\pageref{lastpage}} \pubyear{2011}

\maketitle

\label{firstpage}

\begin{abstract}
MAD is the first multi-conjugated adaptive optics system at the VLT. We present 
{\it H} and $K_S$ observations of the young massive cluster Trumpler\,14 revealing the 
power of MCAO systems by providing a homogeneous Strehl ratio over a large field of view. 
Mean Strehl ratios of 6.0 and 5.9 per cent with maximum Strehl ratios of 9.8 and 12.6 per 
cent in {\it H} and $K_S$, respectively, show significant improvement of the spatial PSF stability compared to single-conjugated adaptive optics systems. Photometry of 
our observations cover a dynamic 
range of $\sim10$ mag including 2--3 times more sources than comparable seeing-limited 
observations. The colour-magnitude diagram reveals that the massive cluster originated 
in a recent starburst-like event $1\pm0.5$ Myr ago. We tentatively detect hints for an 
older population of 3 Myr suggesting that low intensity star 
formation might have been going on in the H{\sc ii} region for a few Myr. We derive the 
luminosity function and mass function between 0.1 M$_{\sun}$ and 3.2 M$_{\sun}$ and  
identify a change of the power law slope of the mass function at 
$m_c\sim0.53^{+0.12}_{-0.10}\ \rm{M}_{\sun}$. The MF appears shallow with power law slopes of 
${\rm\Gamma_{1}=-0.50\pm0.11}$ above $m_c$ and ${\rm \Gamma_{2}=0.63\pm0.32}$ below $m_c$.
\end{abstract}


\section{Introduction}
\begin{table*}
\caption{Strehl values over the 1$\times$1 arcmin FoV} \label{strehls}
\normalsize{
\begin{tabular}{ccccc}
\hline
Origin & Filter & max. Strehl & min. Strehl & mean Strehl\\ \hline
Obs. MCAO & {\it H} & 9.8 \% & 3.3 \% & 6.0 \%\\
Obs. MCAO & $K_S$ & 12.6 \% & 1.3 \% & 5.9 \%\\
Sim. SCAO & $H_{sim}$ & 9.8 \% & $<0.01$ \% & 1.6 \%\\
Sim. SCAO & $K_{S,sim}$ & 12.6 \% & $<0.01$ \% & 2.0 \%\\
\hline
\hline
\end{tabular}
}
\end{table*}
Adaptive optics (AO) systems have proven their outstanding ability in compensating 
atmospheric seeing over more than 20 years. With the first common user AO system, 
ADONIS, mounted at the ESO 3.6m telescope on La Silla \cite{rousset90}, ground-based 
telescopes provide observations with previously unachieved angular resolution. 
However, the performance of a single-conjugated AO (SCAO) system is restricted 
to small fields of view (FoV). With typical isoplanatic angles of 15 arcsec, the 
AO performance degrades quickly at distances of 20--30 arcsec from the guide star (GS). 
Thus, AO assisted observations are limited to maximum field sizes of 
$\sim1$ arcmin with good AO corrections for a typical FoV of 30 arcsec. 
To provide a spatially more stable performance, multi-conjugated AO (MCAO) systems 
are required. MCAO uses several GSs to correct the blurring due to Earth's atmosphere. 
In the framework of second generation instruments for the VLT and the E--ELT, the
Multi-conjugate Adaptive optics Demonstrator (MAD), was developed
in 2007 as the first MCAO system at the VLT (Marchetti et al. 2003, 2004, 2007).

The proximity of the Carina Nebula and its cluster population provides an 
ideal testbed to assess the capability of VLT--MAD. The combination of high 
spatial resolution and the wide field of VLT--MAD allows to resolve 
the dense central regions of the clusters.
NGC\,3372, or Carina Nebula, is a Galactic giant molecular cloud and a site of
vigorous and ongoing star formation including several young massive clusters. It 
includes not only dozens of young O- and B-type stars but also evolved Wolf--Rayet 
stars \citep[e.g.][]{massey93}. The youngest and most populous clusters of the region 
are the open clusters Trumpler\,14 and Trumpler\,16 (hereafter Tr\,14 and Tr\,16), 
located in the central part of the Carina Nebula. The high-mass population 
with its strong UV radiation interacts with the surrounding material triggering 
subsequent star formation \cite[e.g.][]{rathborne02}. The stellar population of Tr\,14 
includes several O-type stars, notably the O2If* star HD93129Aa \citep{walborn} which has
been found to be the most massive star in Tr\,14 with an estimated mass in excess of 100
M$_{\sun}$ \citep{nelan04}. Assuming a Kroupa--IMF, Tr\,14 was suggested to be as
massive as a few $10^3$ M$_{\sun}$ \cite{sana}, which puts it close to the regime of the Galactic 
starburst clusters like NGC\,3603\,YC or the Arches Cluster \cite[e.g.][resp.]{rochau,stolte}. 
The Carina Nebula is the closest region harboring such a massive young cluster, with recent 
studies estimating the distance of Tr\,14 to be between $\sim2$ and $3$ kpc 
\cite[e.g.][]{tapia03,carraro,ascenso07}. 
The lack of evolved stars makes Tr\,14 a very young cluster. The ages of the high-mass 
content of Tr\,14 and Tr\,16 are estimated to be around 1--2 Myr \cite{vazquez96} 
and 2--3 Myr \cite{smith06}, respectively, while the formation of intermediate-mass 
stars probably started earlier. This is supported by the core--halo structure 
of the cluster, with the halo being slightly older \cite{ascenso07}. However, the ages 
of the clusters as well as the distance estimates remain controversial. One reason could be 
the anomalous extinction law towards the clusters \cite[e.g.][]{tapia03}, as this hampers 
precise age determinations, masking uncertainties/differences in the models for low- and 
high-mass stars in the main sequence and pre-main sequence, respectively \cite{naylor}.

We were able to obtain MCAO assisted observations with VLT--MAD to investigate the 
young massive star cluster Tr\,14 (PIs: H. Sana, B. Rochau). In 2010, Sana et al. 
presented their analysis of the data of the entire 2 arcmin$\times$2 arcmin FoV. 
Their scientific focus was on deriving the colour--magnitude diagram (CMD), 
structure of the cluster and to conduct 
a companion analysis. In contrast, our aim is the investigation of the pre-main sequence 
population down to its low- and very-low mass members and the underlying mass 
function. Therefore, we concentrate on the deeper {\it H}- and $K_S$-band observations of 
the central 1 arcmin$^2$ of the cluster. The results of 
our investigation are presented as follows: In Sect.\,2 we introduce shortly 
the MCAO system MAD, followed by the description of the observations and 
the data reduction (Sect.\,3). Sect.\,4 includes a technical analysis of the MCAO performance. 
The scientific analysis is presented in Sect.\,5 and our results are summarised in Sect.\,6.
\begin{figure*}
\centering
\includegraphics[scale=0.5]{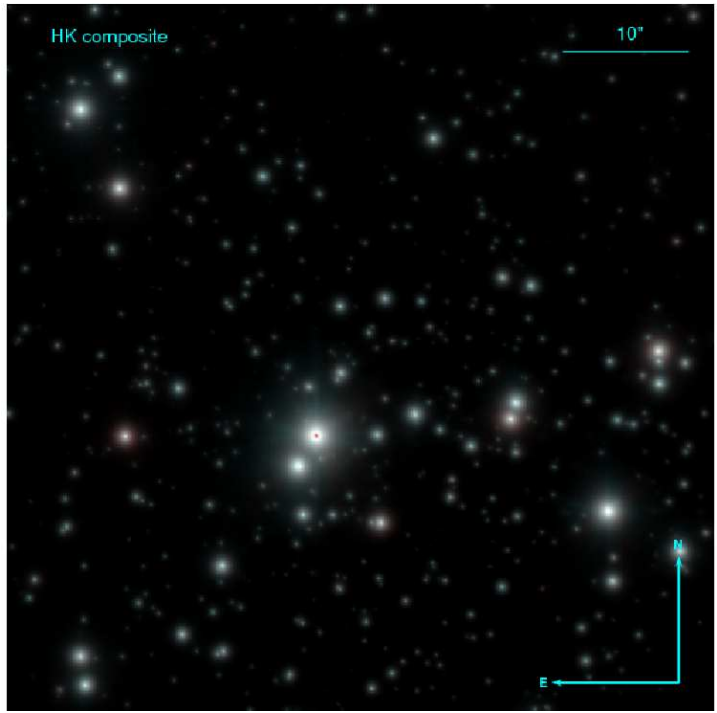}
\caption{Central region of Tr\,14 as seen with VLT--MAD (north is up, east to the left). 
This $HK_S$ colour composite shows the densest region of the cluster including the most 
massive cluster member HD93129Aa $\rm{(\alpha=10^h43^m57.5^s\rm{,}\ \delta=-59^\circ32'52.3")}$ as the brightest star in the field.\label{fig1}}
\end{figure*}

\section{VLT--Multi-conjugate Adaptive optics Demonstrator}
MAD has been developed as the first prototype MCAO system for the VLT. 
It aims at demonstrating the feasibility of MCAO systems in general and to 
investigate different reconstruction techniques in particular.

MAD is designed to correct for blurring due to atmospheric turbulence over 
2 arcmin on the sky using Natural Guide Stars (NGS). Two Deformable Mirrors are
used for the MCAO correction. The first, optically conjugated at the telescope
pupil, corrects for the ground layer turbulence, while the second, conjugated at 
8.5 km elevation above the telescope, corrects for the field anisoplanatism.

To examine the differences between the star- and layer-oriented reconstruction
technique, two different wavefront sensors (WFS) are installed. The star-oriented 
MCAO correction is supported by a Multi Shack--Hartmann WFS. The layer-oriented MCAO 
reconstruction uses a Layer Oriented Multi--Pyramid WFS. Both WFSs can sense simultaneously 
several NGS at visual wavelength but only one mode will be used at a time. Our 
observations were processed in star-oriented mode using 3 NGS.

MAD is equipped with the IR camera CAMCAO (CAmera for MCAO) providing 
near-infrared broad- and narrow-band imaging, using a HAWAII2 2kx2k HgCdTe detector. 
With an image scale of 0.028 arcsec $\rm{pixel^{-1}}$, the camera covers a 59 
arcsec$\times$59 arcsec FoV \citep{amorim}. A 2 arcmin$\times$2 arcmin FoV 
can be covered by moving the camera while the telescope is tracking and the AO 
loop remains closed. In contrast to SCAO, using multiple GSs allows to 
cover a larger AO corrected FoV with homogeneous Strehl ratios.

\begin{figure}
\centering
\includegraphics[scale=0.45]{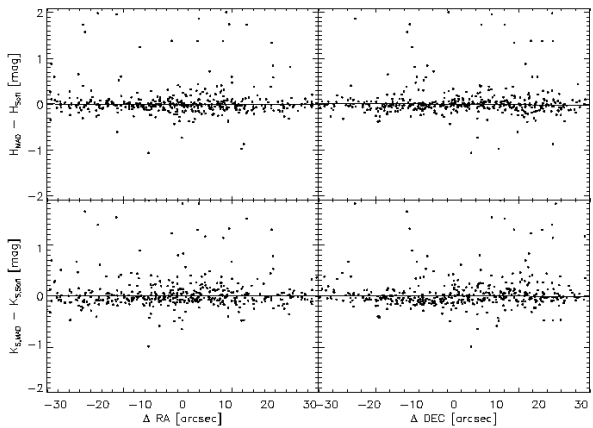}
\caption{Observed differences between the MAD and the NTT-SofI photometry after 
application of the constant ZP correction. ZP offsets are measured and 
plotted as a function of right ascension (left panels) and declination (right panels) 
for the {\it H}-band (upper panels) and $K_S$-band (lower panels). Straight lines show the 
linear fit to the observed distribution revealing the spatial stability of the 
photometry.\label{off}}
\end{figure}
\section{Observations and data reduction}
\subsection{Observations}
Observations were carried out during the night of 2008, January 12 by the VLT--MAD
team. Deep {\it H} and $K_S$ images were taken to map the innermost region of the young 
massive cluster Tr\,14 ($\rm{\alpha=10^h43^m55^s\rm{,}\ \delta=-59^\circ33'03"}$). 
Short integration times of 2 s were used to avoid severe saturation, and 30 
individual integrations are co-added to a single exposure of 1 min. Raw data 
reduction and image combination was performed using the ESO image processing 
software {\it Eclipse} \citep{devillard}. For sky and dark subtraction as well 
as for flat field correction the calibration images taken during the science run are 
used. The sky and dark current subtracted, flat field corrected images are subsequently 
combined using the shift-and-add tool of {\it Eclipse}. 28 single exposures of 1 min in 
each band are combined to a single frame of 28 min of total integration time. For 
comparison, the observations of the adjacent fields provided only 8 min of total 
exposure time (see also Table 2 of Sana et al. 2010).
The final FoV of the central field covers an area of 68 arcsec$\times$68 arcsec, 
and the $HK_S$ colour composite is shown in Fig.\,\ref{fig1}. Located close to the centre 
is HD93129Aa, the most massive member and brightest star in the field.

\begin{figure*}
\centering
\includegraphics[scale=0.725]{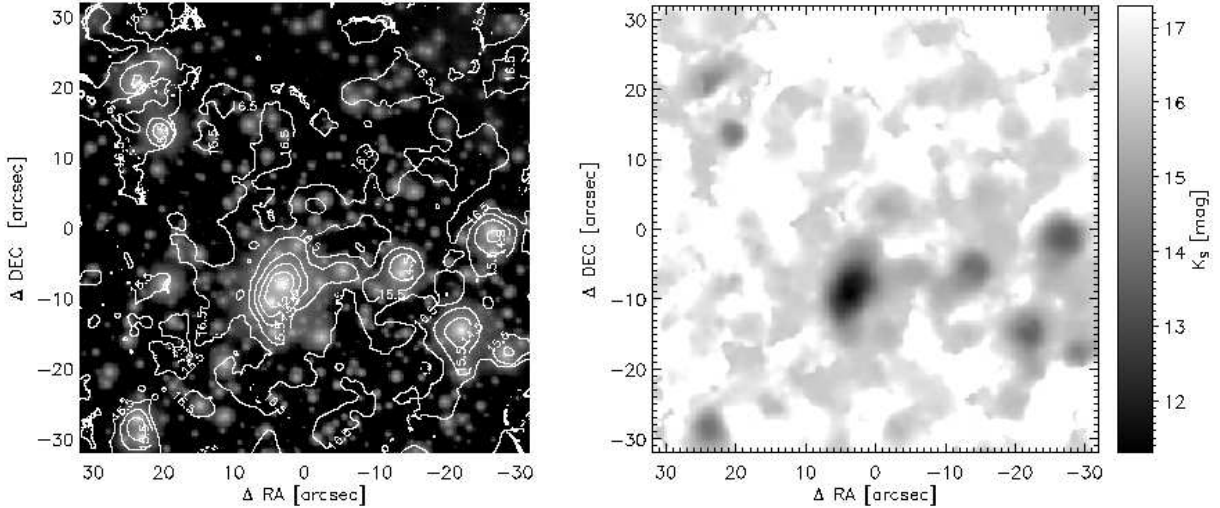}
\caption{$K_S$-band completeness map of Tr\,14. {\it Left Panel:} $K_S$-band image 
of Tr\,14, north is up and east to the left. Magnitudes at which 70 per cent 
completeness is achieved are superimposed as white contours (in steps of 1 mag). 
The image is centered (0/0) at $\rm{\alpha=10^h43^m57.2^s\rm{,}\ \delta=-59^\circ32'44.2"}$. 
{\it Right Panel:} $K_S$-band completeness map of the entire observed FoV. The grey 
shades correspond to the $K_S$-band magnitude at which a completeness of 70 per cent 
is achieved. The strong local decrease of completeness close to bright stars without 
displaying large scale variations across the field shows that crowding 
effects are less important compared to the possible faintness of the stars.\label{cmpl}}
\end{figure*}

\subsection{Photometry}\label{daoph}
To study the stellar content of Tr\,14, we perform PSF photometry using
{\sc IRAF/DAOPHOT} \citep{stetson}. Fitting 15 stars in {\it H}- and $K_S$-band, 
we find a Penny-PSF, comprising a Gaussian kernel and Lorentzian wings, 
as the best-fitting PSF. It further turns out that a second order variable 
PSF provides a good match to the MAD data obtained with the star-oriented 
mode. Photometric calibration is based on NTT--SofI observations of Tr\,14 published 
by Ascenso et al. (2007). The detectors and filter systems of MAD and NTT--SofI 
are similar, leading to the absence of e.g. colour terms. Consequently, using 
the photometric catalogue of the MAD and NTT--SofI observations is well suited to 
assess the goodness of the PSF characterization of our MCAO observations.

\begin{figure}
\centering
\includegraphics[scale=0.45]{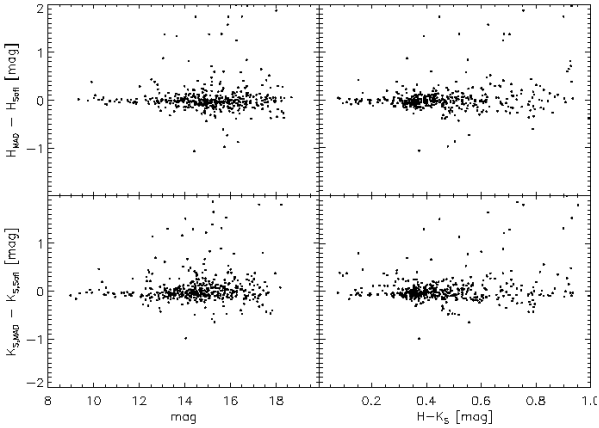}
\caption{Observed differences between the MAD and the NTT-SofI photometry after 
application of the constant ZP correction as a function of magnitude (left panels) 
and colour (right panels) for the {\it H}-band (upper panels) and $K_S$-band 
(lower panels).\label{offb}}
\end{figure}

The comparison of seeing-limited NTT-SofI photometry with our MAD photometry 
reveals no trend of photometric zeropoint (ZP) with position on the 
MAD field ($\sim$1.1 arcmin$\times$1.1 arcmin) and shows a rms scatter of 0.11 mag 
in $H$-band and 0.12 mag in $K_S$-band, respectively (Fig.~\ref{off}). Furthermore, 
Fig.~\ref{offb} displays the ZP offset to be also constant with magnitude 
and colour. This agrees very well with the results of Sana et al. (2010) 
who applied PSF photometry to individual frames of 1 min instead of the combined 
image as used in our analysis. The spatially stable photometry emphasises 
that the stability of the PSF is comparable to that of seeing-limited observations, and 
exceeds the stability of SCAO imaging, where anisoplanatic effects result in considerably 
larger variations in the PSF across the FoV.

Thus, the combination of star-oriented MAD observations with a second order variable 
PSF yields high precision photometry across the entire FoV.
The final photometric catalogue includes 1347 stars. Excluding stars with {\sc DAOPHOT} 
photometric uncertainties above 0.1 mag (375 sources), we used a catalogue of 972 
stars for the further analysis comprising stars over a dynamic range of $\sim10$ 
mag. Comparison to the 453 sources detected with NTT-SofI in the 
MAD field shows that VLT--MAD was able to detect 2-3 times as many sources.

\begin{figure*}
\centering
\includegraphics[scale=0.725]{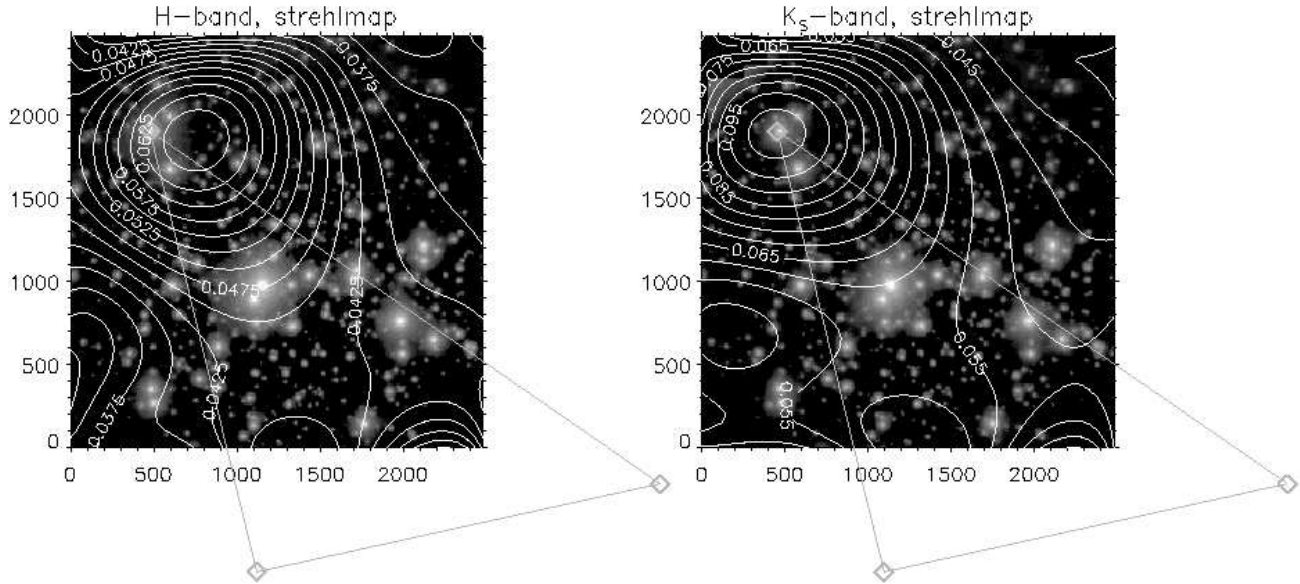}
\caption{H- and $K_S$-band image of the core of Tr\,14 (left and right panel, 
respectively). North is up and east is to the left. Overplotted contours show the 
level of equal Strehl ratios in {\it H}- and $K_S$-band. Apparent is the smooth 
decrease of Strehl values with the peak value centered on the GS included in the 
field in the $K_S$-band. The offset of the peak Strehl to the position of the GS 
in the {\it H}-band is an artefact caused by the saturation of the star. Positions 
of the GSs are indicated by the grey diamonds and the encircled triangle 
by the straight grey lines.\label{fig2}}
\end{figure*}

\subsection{Completeness}\label{compl}
We aim at deriving the luminosity function (LF) and, 
subsequently, the mass function (MF) of Tr\,14. To derive the correct cluster 
LF and MF, accurate knowledge of the completeness of the photometric catalogue 
is required. In dense stellar fields the detectability of a source depends on 
its flux and the local stellar density ('crowding'), and thus, on brightness and position 
of the source in the observed field. Crowding describes the effect of a 
decreasing detection probability due to high stellar densities in the sense that 
a high brightness contrast between a point source and a possible 
bright neighbour limits its detectability.

To determine the completeness, we apply the same technique as described in 
Gennaro et al. (2011). We will briefly summarise the procedure here and 
refer to the Appendix of Gennaro et al. (2011) for further details. 
Artificial stars are inserted at random positions in the field, and the 
data analysis is re-run to estimate the fraction of recoveries. We added 50 
artificial stars per run to our science image using the PSF derived with 
{\sc DAOPHOT}. The rather low number of artificial stars has been chosen in order 
to not change the crowding characteristics of the observations. 100 individual 
runs are processed to add a total of 5,000 stars to our image in each photometric 
bin of 0.5 mag width. We achieve a typical separation between simulated stars of 
$d_{sim}\sim34\ \rm{pixel}$.

The final product of our procedure for each photometric band 'j' is the 
completeness as a function of position on the detector and magnitude of the star:
\begin{equation}
{\rm C_j({\it x,y,\mu})=\frac{\alpha({\it x,y})}{exp({\frac{\mu-\beta({\it x,y})}{\gamma({\it x,y})})+1}}}
\end{equation}
Here $\mu$ is the magnitude of the corresponding star, 
${\rm \alpha}$ ($\leq1$) is the normalization factor, ${\rm \beta}$ the magnitude 
at which the completeness is ${\rm \alpha/2}$, and ${\rm \gamma}$ describes how fast 
the completeness ${\rm C_j}$ drops to zero with decreasing brightness. With the derived 
values for ${\rm (\alpha,\beta,\gamma)}$, we can assign a completeness factor to each 
star considering its magnitude and position.

Fig.~\ref{cmpl} shows the $K_S$-band image of Tr\,14 with the 
70 per cent completeness limits at different magnitudes superimposed 
as contours (left panel). The right panel depicts the same 
completeness limits over the field as a function of 
magnitude. The limiting magnitude is significantly brighter in 
the proximity of the brightest stars. In less dense areas with 
fainter stars the completeness shows a smooth distribution 
over the field with a rather faint average limiting magnitude. 
In addition with the proximity of the cluster, this reveals 
that crowding is not a major source of incompleteness compared 
to the stellar faintness. Together with the largely increased number 
of detections when compared to the seeing-limited NTT--SofI observations, 
this illustrates the improved performance provided by MAD.

\begin{figure*}
\centering
\includegraphics[scale=0.6]{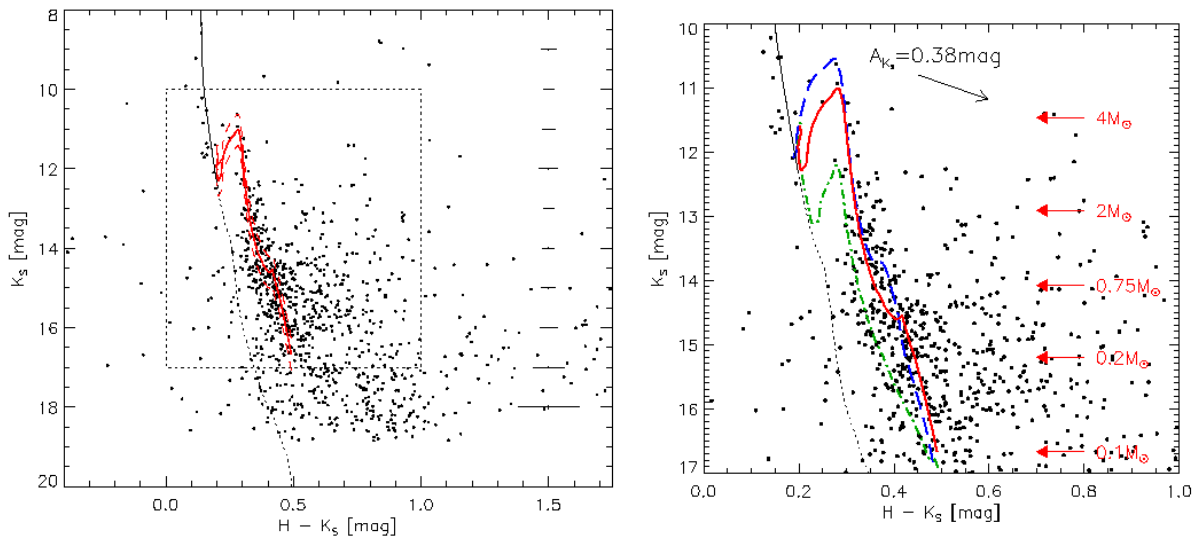}
\caption{$K_S$ vs. $H-K_S$ CMD of the central 0.75 pc of Tr\,14. The cluster population 
can be identified as the well populated PMS in the CMD and goes down to (and beyond) 
the H-burning limit. {\it Left Panel:} The CMD is well represented by a Siess isochrone 
of 1 Myr in age (red solid line); red dashed lines show the uncertainties in the distance 
modulus. The MS isochrone is shown by the thin and the dotted black line. The box (dotted 
lines) displays the region that is shown in the right panel. {\it Right Panel:} 
The CMD with the 0.5 Myr, 1 Myr and 3 Myr Siess isochrones overplotted (blue dashed, red solid 
and green dash-dotted line, respectively). Labelled masses are taken from the 1 Myr Siess isochrone.\label{cmd}}
\end{figure*}

\section{Technical analysis}
The observations have been carried out under observing conditions with a typical 
seeing between 0.7 and 1.6 arcsec. The MCAO correction provide data with 
an average full width at half maximum (FWHM) of 0.22 and 0.26 arcsec 
in {\it H}- and $K_S$-band, respectively. The relatively bad seeing conditions, under 
which the observation were accomplished, hampered a better correction. Considering 
the good geometry and brightness of the GSs together with good seeing conditions 
allow MAD to easily achieve AO corrections down to a FWHM of 0.15 arcsec, or 
even below \cite[e.g.][]{momany,campbell}. Seeing-limited NTT--SofI observations of 
Tr\,14 by Ascenso et al. (2007) had, for comparison, a FWHM around 0.7 arcsec. 

To investigate the performance of the MCAO system we measure the Strehl 
ratio of the combined image across the field. It is defined as 
the ratio of the measured peak intensity to the peak intensity of an ideal 
diffraction-limited image of a star of equal magnitude. Stellar properties are 
extracted from the combined images using the Source Extractor software 
\cite{bertin}. The theoretical diffraction-limited PSF is created with 
the {\it imgen} task in the {\it Eclipse} package, using a primary and secondary 
mirror aperture of 8.2 and 1.1 m, respectively, a central wavelengths of 
$\lambda_H=1.65\ \mu m$ and $\lambda_{K_S}=2.125\ \mu m$, with a width of 
$\Delta \lambda_H=0.3\ \mu m$ and $\Delta \lambda_{K_S}=0.35\ \mu m$. We derive 
the Strehl ratios of our {\it H}- and $K_S$-band observations and the distributions 
of Strehl ratios in $H$- and $K_S$-band over the observed FoV are shown in 
Fig.\,\ref{fig2} as white contours. As expected in the 
star-oriented mode of MAD, the Strehl ratio is maximised on the GS. Maximum Strehl 
values of 9.8 per cent ({\it H}-band) and 12.6 per cent ($K_S$-band) are measured. 
The Strehl map, furthermore, reveals a shallow decrease of the Strehl ratio over the 
FoV. This is indicated by mean Strehl ratios of 6.0 per cent for {\it H}-band and 5.9 
per cent for $K_S$-band. The results are summarised in Table\,\ref{strehls}. 
A steep decrease of the AO performance, as observed in SCAO observations, 
is alleviated due to the decreasing distance towards the other two GSs which 
are located outside our FoV.

To quantify the difference between MCAO and SCAO, we compare our observations to 
simulated  SCAO observations. We create a grid representing our observed MAD FoV 
and assign values of Strehl ratios to each grid element with maximum Strehl ratios 
at the GS location according to our observations. 
The spatial variation of the Strehl ratio is assumed to follow 
${\rm SR=S_0exp(-[\frac{\Delta \Theta}{\Theta_{iso}}]^{5/3})}$ \citep{cresci05} with 
${\rm S_0}$ as the maximum Strehl ratio and an isoplanatic angle ${\rm \Theta_{iso}=15}$ 
arcsec as a typical value for SCAO observations. Results of the simulations are also 
summarised in Table\,\ref{strehls}. We found mean Strehl values of 1.6 and 2.0 per cent 
for the {\it H}- and $K_S$-band, respectively. 
The larger mean Strehls in our MAD observations reveal the improved spatial stability of the 
PSF in our MCAO observations compared to SCAO imaging.

Although SCAO observations provide higher peak Strehl ratios, the fastly decreasing 
performance with distance to the GS makes a good characterization of the PSF at larger 
distances difficult. Especially in crowding-limited regions, such as dense stellar clusters, 
this hampers the census of the stellar population. The greater homogeneity of the PSF 
provided by MCAO observations allows, therefore, the coverage of larger AO corrected areas 
on the sky, crucial for photometric and astrometric studies of extended stellar populations.

\begin{figure*}
\centering
\includegraphics[scale=0.6]{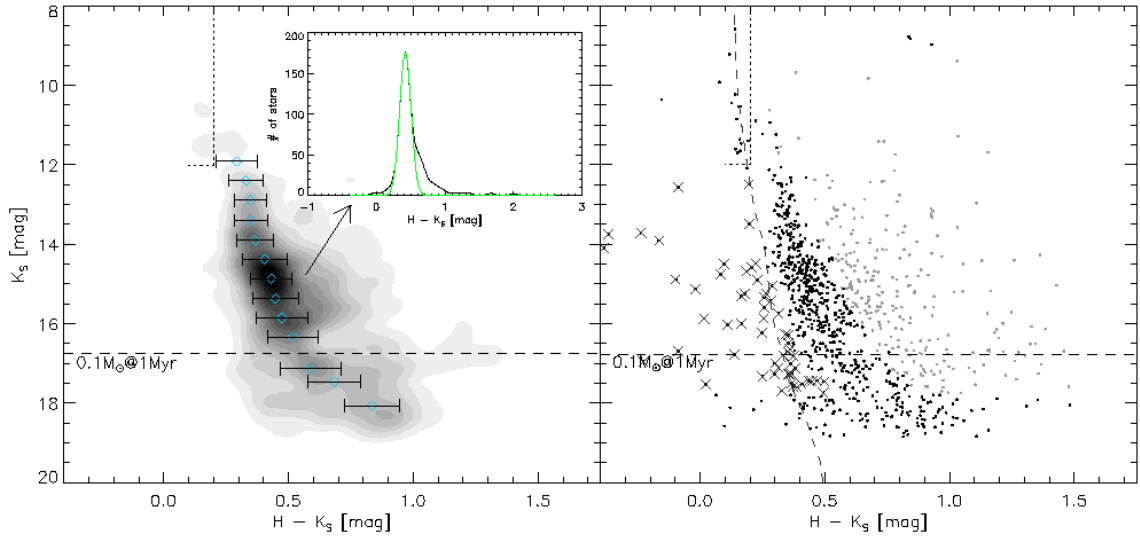}
\caption{{\it Left Panel:} Hess diagram based on our photometric 
catalogue. To distinguish the PMS stars without infrared excess 
from the contaminating field population and excess sources in 
the CMD we cut the Hess diagram into stripes of 0.5 mag in magnitude. 
Blue diamonds show the central magnitude of each stripe (y-axis) and the colour 
of the peak of the underlying stellar density (x-axis). The errors bars 
show the 1$\sigma$-deviation from the fitted Gaussian function. PMS 
stars without significant infrared excess are identified by being inside 
the 1.5$\sigma$ limit. {\it Inlay Panel:} An example of the density distribution 
for one magnitude bin (here: 14.6 mag$<K_S<15.1$ mag). The black line shows the 
stellar density as a function of colour. A Gaussian function is fitted around the 
peak of the distribution and overplotted as the green line. {\it Right Panel:} 
The CMD as in Fig.~\ref{cmd} with the identified field contaminants 
and 'excess' sources highlighted 
(crosses and grey dots, respectively). The field contamination has been excluded 
while deriving the LF of Tr\,14 while the 'excess' population is considered. 
Dotted lines mark the region of the excluded MS population. The dashed line 
is the 1 Myr MS isochrone by Marigo et al. (2008) and nicely displays the 
good identification of the field contaminants.}
\label{hess}
\end{figure*}

\section{Scientific analysis}\label{science}
\subsection{Colour--magnitude diagram}\label{cmd_section}
Using the photometry of Sect.\,\ref{daoph}, we create a $K_S$ vs. $H-K_S$ CMD, 
shown in Fig.\,\ref{cmd}. The CMD allows the identification of the 
cluster population, primarily the cluster pre-main sequence (PMS) population 
between 10 mag$<K_S<$16 mag and 0.2 mag$<H-K_S<$0.5 mag. The transition 
region from PMS to the main sequence (MS) is sparsely populated but can 
be identified at 10.5 mag$<K_S<12.5$ mag with $0.1$ mag $<H-K_S<0.3$ mag. 
Stars redder than the PMS ($H-K_S\gtrsim$0.5 mag) suffer either from severe differential extinction 
in the region, are background objects, highly reddened by the Carina Nebula or 
belong to the cluster as PMS stars surrounded by circumstellar material such as 
a disc causing the $K_S$-band excess.

Without a comparison field, we cannot precisely quantify the contaminating 
fore- and background field star population and, hence, subtract statistically 
the contaminating field. However, the identification of different features of 
the cluster in the CMD (PMS, PMS-MS transition region) still allows the 
comparison with stellar evolutionary models. For the PMS, we use Siess tracks \citep{siess00}, 
computed as in Da Rio et al. (2009), using the BT--Settl Grid of Allard et al. (2010).
MS isochrones are from Marigo et al. (2008). Isochrones that resemble best the 
observed features are shown in the right panel of Fig.\,\ref{cmd} superimposed onto the CMD. 
The shape and width of the PMS-MS transition has proven to be a good age 
estimator \cite[e.g.][]{brandner,rochau}, but in the case of Tr14 the region is 
sparsely populated. Hence, to estimate the age of the cluster we use the colour 
difference between the MS and the PMS as well as the shape of the PMS. We identify 
the cluster PMS as best represented by the 1 Myr isochrone indicating that Tr\,14 is 
a very young stellar population. The derived distance modulus ($DM=11.8\pm0.4$ mag) 
corresponds to a distance of $\sim2.3\pm0.4$ kpc. The uncertainty is estimated by 
finding the closest and farthest distance which gives an acceptable approximation of the 
PMS of the cluster. The extinction was simultaneously derived to $A_{K_S}=0.38\pm0.03$ mag, 
using $R_V=4.16$ \cite{carraro} and the relation of visual and near-infrared extinction 
according to Cardelli et al. (1989). The error on the average extinction 
is estimated by shifting the isochrone towards the blue and red boundary of the upper 
PMS, respectively.

The distance of 2.3 kpc is comparable, though slightly smaller, to earlier 
derived distance estimates. For example, Carraro et al. (2004) locates Tr\,14 at a distance 
of 2.5 kpc, while Tapia et al. (2003) estimated a somewhat larger distance of 2.63 kpc. 
Comparing our extinction estimate, which corresponds to an $A_V=3.0\pm0.4$ mag following 
the applied extinction law, reveals a slightly higher average extinction towards Tr\,14 
compared to earlier values, e.g. $A_V=2.6\pm0.2$ mag \cite{tapia03} or $A_V=2.0\pm0.2$ 
mag \cite{carraro}. We benefit from a higher angular resolution, and the improved photometry 
might explain the slight differences observed in the different studies.

The best fitting values, distance$=2.3\pm0.4$ kpc and $A_{K_S}=0.38\pm0.03$ mag, are derived using 
a 1 Myr PMS isochrone. This isochrone fits the PMS-MS colour width and shape of the PMS best.
With a different combination of distance and extinction 
(distance$=2.6\pm0.4$ kpc, $A_{K_S}=0.36\pm0.03$ mag), a slightly younger isochrone 
(0.5 Myr) represents the PMS-MS colour difference well, but the PMS shape is less well 
reproduced (Fig.\,\ref{cmd}). For the 2 Myr isochrone the colour difference between MS 
and PMS is very narrow compared to the observations. We therefore adopt an uncertainty 
of 0.5 Myr on the age. However, very young stellar ages should be dealt with carefully 
as evolutionary models have large uncertainties for ages $\lesssim 1$ Myr, given the strong 
dependence of the predicted luminosities and temperatures on the initial conditions of 
the computation \cite[e.g.][]{baraffe}.

Comparing the CMD with the Siess isochrones, we find the 1 Myr isochrone to best 
resemble the PMS. However, at 14 mag$<K_S<16$ mag we identify a number 
of stars with colours bluer than the isochrone. 
Apart from observational biases such as photometric uncertainties or a possible lower 
extinction of the identified sources compared to the derived average extinction, 
the observed sequence could alternatively be explained by a somewhat older population 
of PMS stars. By superposition, we observe that a 3 Myr isochrone nicely follows this 
blue sequence (see Fig.\,\ref{cmd}). An older population would support earlier 
statements of continuous star formation in the Carina Nebula, creating a surrounding
older halo population with ages up to 5 Myr \citep{ascenso07}.

\begin{figure*}
\centering
\includegraphics[scale=0.85]{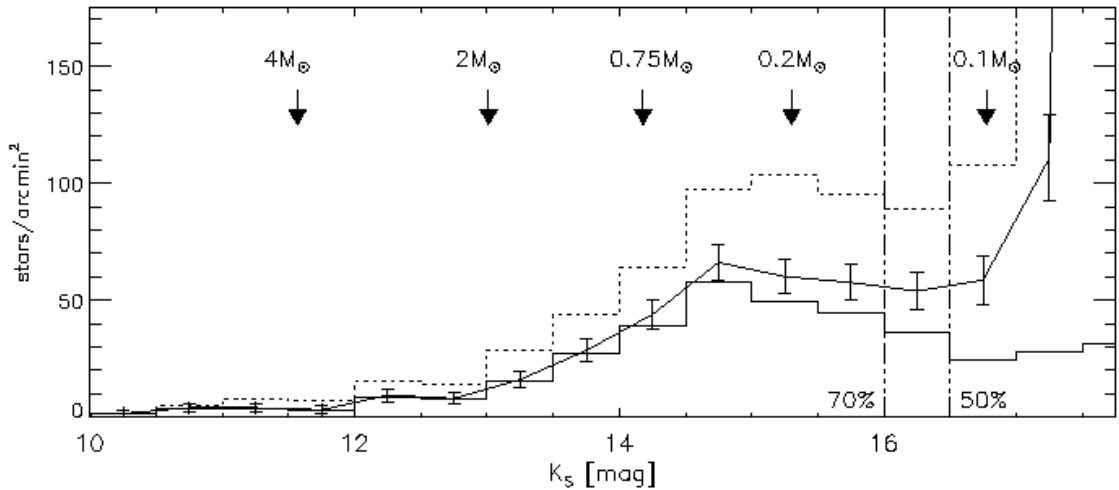}
\caption{$K_S$-band LF of Tr\,14. Displayed 
is the number of stars found in magnitude bins of 0.5 mag normalised to the 
observed area. The histogram shows the number of detected sources while 
the solid line depicts the same number but corrected for incompleteness. 
The dashed histogram is the completeness corrected LF including also 'excess' sources. 
The errors are the Poisson errors of each bin. Corresponding masses, highlighted 
by the arrows are taken from the 1 Myr Siess isochrone. The LF shows an increasing number 
of stars with decreasing magnitude to a turnover observed in the 
$0.25-0.54\ \rm{M}_{\sun}$ bin (PMS) and/or in the 
$0.18-0.25\ \rm{M}_{\sun}$ bin (PMS+'excess' sources). The dash-dotted lines 
mark the limit at which the completeness reaches 50 per cent and 70 per cent, respectively.}
\label{klf}
\end{figure*}

\subsection{Luminosity function}\label{lf}
\subsubsection{Selection of different populations}\label{sel}
Since the observations have been carried out without observing a 
comparison field, the isolation of the cluster population remained 
challenging. In order to get a better idea of the cluster population 
we apply the following approach: We create a Hess diagram based 
on our {\it H}- and $K_S$-band observations (left panel of Fig.~\ref{hess}) which 
shows the stellar density in the colour--magnitude plane rather than the 
single stars.

The Hess diagram is divided into 20 stripes, 0.5 mag wide. In each stripe 
we examine the stellar density as a function of colour.
An example of such a distribution is shown in the inlay diagram in 
Fig.~\ref{hess}. Around each peak we find a rather symmetric distribution which is 
well represented by a Gaussian function (peak positions are highlighted 
as blue diamonds in Fig.~\ref{hess}). The best-fitting Gaussian function is 
shown by the green line superimposed to the stellar density distribution. 
1$\sigma$ deviations from the peak are plotted as horizontal error bars. 
The sparsely populated cluster MS yields unreliable fits. Hence, we 
excluded sources with $K_S<12$ mag and $H-K_S<0.2$ mag (dotted lines in Fig.~\ref{hess}). 
This does not change our results, 
as we intend to focus in the present work on the PMS population. Thus, we aim to 
isolate the PMS stars without infrared excess from lower mass MS stars belonging 
to the region but not to the young cluster and foreground sources, for simplicity 
in the following together called field contamination, as well as from the sources 
with $K_S$-band excess. Comparison of the Hess diagram with the CMD reveals that a 
limiting deviation of 1.5$\sigma$ divides best the different populations. 
Sources being off the peak by more than 1.5$\sigma$ are considered as field 
contamination (bluer than peak) or as 'excess' cluster sources 
(redder than the peak) and displayed by the crosses and grey dots in the right 
panel of Fig.~\ref{hess}, respectively. The limit has been chosen such that stars 
which can be clearly assigned to the contaminating field are separated from the PMS of the 
cluster. The PMS is nicely confined along a relatively narrow sequence and the chosen 
limit provides a good division of the PMS sources into a non-excess and an excess sequence.

Although we cannot clearly identify the origin of the observed 
excess, these stars are considered as excess sources of the cluster. Firstly, the 
cloud of the Carina Nebula is effectively blocking contaminating background objects. 
Furthermore, the effect of differential extinction should be lowered by the presence 
of the massive stars which removed the gas remaining from the star formation process 
such that highly reddened sources are unlikely to suffer from a large extinction. 
We identified 79 sources with infrared excess and 274 PMS stars without 
excess and, assuming all these stars to belong to the cluster, the fraction of excess 
sources would only amount to $\sim22.5$\%, which is lower than the excess fraction 
of 35\% as reported by Ascenso et al. (2007). 
Constraining the different regions of the CMD is in our case the best way to 
get a least biased view of the stellar population of Tr\,14 and, 
subsequently, the best cluster LF.

\subsubsection{$K_S$-band Luminosity Function}
The LF is the number of stars as a function of magnitude. Stars are 
counted in magnitude bins of 0.5 mag, the counts are corrected for incompleteness 
and normalised to the considered area. The final $K_S$-band LF is shown in 
Fig.~\ref{klf}. The histogram displays the LF based 
on stars that belong to the PMS stars without excess (black dots in Fig.~\ref{hess}) 
and is not yet corrected for incompleteness. The completeness corrected 
LF is shown by the straight line with error bars that correspond to Poisson errors. 
The steep increase for stars fainter than $K_S\sim16.5$ mag in the completeness 
corrected LF is due to the high incompleteness at the faint end. The dotted line 
further displays the LF of PMS stars including excess sources (black and grey 
dots in Fig.~\ref{hess}) and is corrected for incompleteness.

The LFs depict an increasing number of stars towards fainter magnitudes at the 
bright end until it turns over at $K_S=14.75$ mag (PMS--LF) or $K_S=15.25$ mag 
('PMS+excess'--LF), respectively. This same turnover is also visible at 
$K_S\sim15$ mag in the two $K_S$--LFs derived by Ascenso et al. (2007) based on 
NTT--SofI and VLT--NaCo observations, respectively.

Similar $K$-band LFs have been identified in clusters, observationally
and by simulating model LFs of synthetic clusters \cite[e.g.][]{zinn+mcc,ladalada}. 
Zinnecker \& McCaughrean (1991) simulated primarily $K$-band LFs for 
clusters of non-accreting PMS stars. They found time-dependent LFs exhibiting maxima 
that shift with time. Comparison of our LF with its maximum at $K_S\sim14.75$ mag 
with the results of Zinnecker \& McCaughrean (1991) reveal comparable functions 
for clusters between 0.7 and 1 Myr, consistent with the derived age for Tr\,14. 
Such maxima are most likely caused by Deuterium-burning 
stars that lead to a peak in the LF. In a 1 Myr old cluster a 
PMS star of 0.3 M$_{\sun}$ starts burning Deuterium, in agreement with the 
masses that correspond to the observed peak in the LF.

\subsection{Mass function}\label{mf}
\subsubsection{The Trumpler\,14 Mass Function}
The initial mass function (IMF) represents the distribution of stellar masses for 
a given stellar system at the time of its formation. The form of the MF
changes with time due to stellar, as well as dynamical evolution. As a result of such 
evolutionary effects, which affect entire stellar systems such as star clusters 
\citep{baumgardt}, the MF is changed and, thus, the 'actual' MF will be called 
present-day mass function (PDMF). 
The PDMF is described, similar to the IMF, as a power law:
\begin{equation}
{\rm \Gamma_{PD}=\frac{d\log{N(\log{\rm m})}}{d\log{\rm m}}\ {,}}
\label{alpha}
\end{equation}
where $N(\log{\rm m})$ is the PDMF. This slope is given by the linear
relation between the logarithmic mass intervals and the corresponding
number of stars (in logarithmic scale).
With this notation the Salpeter-IMF has a slope of $\Gamma=-1.35$ \cite{salpeter}.
Due to the youth of Tr\,14, most of its members are populating the PMS which 
ends with the ignition of hydrogen in stellar cores at the MS turn-on. To obtain 
the mass of all PMS stars, we use the relation of the stellar luminosity 
and mass, the mass--luminosity relation (MLR), of the 1 Myr Siess isochrone. 
To convert apparent to absolute magnitudes we consider the distance of $2.3\pm0.4$ kpc 
and a constant foreground extinction of $A_{K_S}=0.38\pm0.03$ mag as derived in Sect.\,\ref{cmd_section}. 
The mass of each star is estimated with the derived MLR and the 
corresponding magnitude. In case of the PMS stars without excess we have considered the 
brightness of the stars in $K_S$-band, as this minimises the effect of differential 
extinction. PMS stars with circumstellar material are, however, exhibiting an excess in the 
$K_S$-band. Therefore, we have used the $H$-band instead to reduce the effect of the excess 
when deriving the mass.

\begin{figure*}
\centering
\includegraphics[scale=0.75]{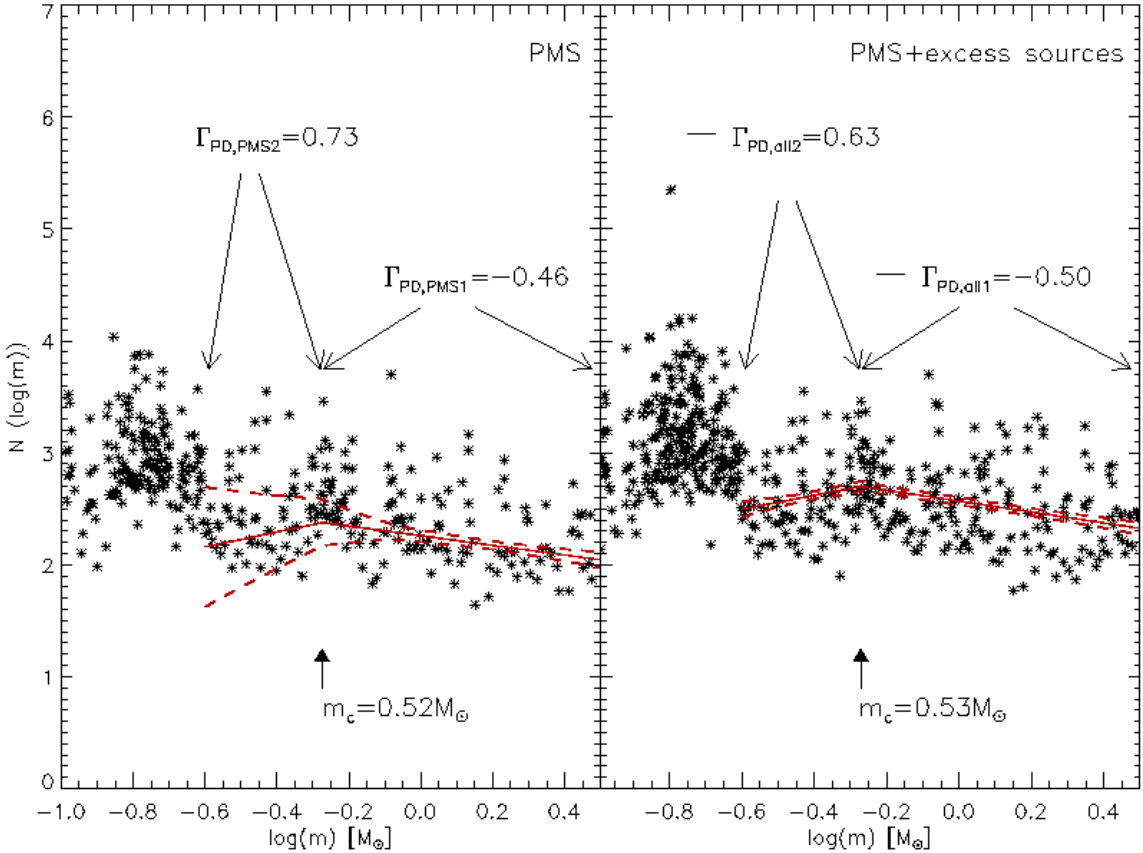}
\caption{MF of Tr\,14 derived following the suggested method of 
Ma{\'{\i}}z Apell{\'a}niz \& {\'U}beda (2005). Each 'bin' contains a single star, which has 
been corrected for incompleteness and is divided by the width of each bin. The two 
MFs are composed of stars of only the PMS (left panel) and of the PMS plus the 'excess' 
sources (right panel). The best fits to the MFs are shown as straight red lines while 
dashed lines display the 1$\sigma$ interval. The 'high-mass' part is well represented 
by a power law with a slope of ${\rm \Gamma_{PD,PMS1}=-0.46\pm0.20}$ (PMS) and 
${\rm \Gamma_{PD,all1}=-0.50\pm0.11}$ (PMS+'excess' sources). The 'intermediate' part appears increasing 
with stellar mass (${\rm \Gamma_{PD,PMS2}=0.73\pm0.59}$ (PMS) and 
${\rm \Gamma_{PD,all2}=0.63\pm0.32}$ 
(PMS+'excess' sources) with a turnover observed of $m_c=0.52^{+0.25}_{-0.17}\ \rm{M}_{\sun}$ (PMS) 
and $m_c=0.53^{+0.12}_{-0.10}\ \rm{M}_{\sun}$ (PMS+'excess' sources).}
\label{kmf}
\end{figure*}

To obtain the PDMF of Tr\,14, we use the method 
suggested by Ma{\'{\i}}z Apell{\'a}niz \& {\'U}beda (2005). 
They discuss the biases that are introduced by the binning 
process to derive the MF \cite[see also][]{maiz09}. Assuming a 
constant bin size can lead to misleading results, due 
to the correlation between the number of stars per bin 
(higher for lower masses) and the assigned weights 
(from the Poisson statistics). To circumvent this problem 
they suggested a variable bin size to have an equal 
number of sources in each bin. They found, in this case, 
that the bias is almost independent on the assumed number of stars 
per bin, and remains low even for a single star in each bin.

To minimise the effect of the binning and to maintain the 
statistical information, we have chosen to derive the PDMF 
in single star bins. Therefore, we sort the stars of our 
catalogue by mass. Each 'bin' consists of a single star and 
its bin width is defined to be the mean of the star and the preceding 
and following star, respectively. For the most and least massive stars in the 
sample, the upper and lower boundaries of the bins are chosen 
to be symmetric from the actual mass of the star. The 'number' (1 by definition) 
of stars is corrected for incompleteness associated with the star 
and divided by the width of the bin.

The resulting PDMF is illustrated in Fig.~\ref{kmf}. It displays two functions 
composed of {\it i)} only the PMS stars (index: PMS$_i$) and of {\it ii)} PMS stars 
and the 'excess' sources (index: all$_i$). In both cases, we identify a PDMF with 
a turnover at $log(m)\sim-0.25$. The jump at $log(m)\sim-0.6$ is due to a 
significant change in the slope of the MLR ($\sim 0.06\ \rm{M}_{\sun}$/mag for 
$m<0.25\ \rm{M}_{\sun}$ and $\sim0.76\ \rm{M}_{\sun}$/mag for $m>0.25\ \rm{M}_{\sun}$). 
Therefore, and due to a probably increasing contribution of the Galactic field 
stars, the PDMF below $log(m)<-0.6$ is not considered in the following 
fitting process. Thus, we fit a two-component power law with a variable characteristic 
mass in the mass range $0.25<\rm{M}<3.2\rm{M}_{\sun}$.

{\it i) PDMF of PMS stars without excess sources (black dots in Fig.\,\ref{hess}):}

We find a characteristic mass of $m_c=0.52^{+0.25}_{-0.17}\ \rm{M}_{\sun}$ 
and power law slopes of ${\rm \Gamma_{PD,PMS1}=-0.46\pm0.20}$ above $m_c$ and 
${\rm \Gamma_{PD,PMS2}=0.73\pm0.59}$ below $m_c$. 

{\it ii) PDMF of PMS stars with excess sources (black and grey dots in Fig.\,\ref{hess}):}

We derived power law slopes of ${\rm \Gamma_{PD,all1}=-0.50\pm0.11}$ and ${\rm \Gamma_{PD,all2}=0.63\pm0.32}$ 
above and below a turnover mass of $m_c=0.53^{+0.12}_{-0.10}\ \rm{M}_{\sun}$, respectively. 
The slopes in both regimes agree very well within the fitting uncertainties. 
Consequently, no systematic bias is introduced due to the inclusion of excess sources and 
the derivation of the PDMF slope is independent of our cluster member selection. Therefore, when 
addressing the PDMF of Tr\,14, we refer in the following to the PDMF including the excess sources, i.e. 
to ${\rm \Gamma_{PD,all1}=\Gamma_1=-0.50\pm0.11}$ and ${\rm \Gamma_{PD,all2}=\Gamma_2=0.63\pm0.32}$ 
above and below $m_c=0.53^{+0.12}_{+0.10}\ \rm{M}_{\sun}$.

The turnover mass that is identified agrees very well with the mass at which the  
flattening of the Kroupa--IMF occurs \cite{kroupa}. Such flattening of the MF 
around $\sim0.53\ \rm{M}_{\sun}$ is not an unique feature of Tr\,14, but observed in other 
clusters \cite[e.g. Fig.\,3 in][]{bastian}.

The slopes of the PDMF in the different mass regimes (above and below $m_c$, respectively) 
are notably shallower than the corresponding part of the Kroupa--IMF (Salpeter slope $\Gamma=-1.35$). 
However, a source of uncertainty in the derivation of the MF slope are 
unresolved binaries. Depending on the binary fraction and the intrinsic slope of the stellar 
MF, the observed slope may deviate from the intrinsic slope by up to $\Delta \Gamma \sim+0.5$ 
\cite[e.g.][]{sagar,kroupa}. Even with such a large correction our PDMF would appear flatter 
than a Kroupa--IMF.

Considering the rather flat slope of the PDMF of the central region of Tr\,14 and the 
unaccounted effect of unresolved binaries, the derived PDMF is in good agreement with the 
central PDMFs observed in other young and massive clusters. In  the very central 
region in NGC\,3603\,YC slopes of the MF are found between $\Gamma\sim-0.31$ 
and -0.91, depending on the stellar population (MS, PMS, PMS plus MS) 
and area considered in deriving the MF \cite[e.g.][]{sung,stolte06,harayama}. The different 
considered mass ranges do not hamper the comparison as above 0.5 M$_{\sun}$ the 
MF is described by a single-power law \cite{salpeter,kroupa}.
In the case of Westerlund\,1, Gennaro et al. (2011) found a significant flattening of the PDMF towards 
the centre of the cluster with slopes down to $\Gamma\sim-0.7$ although the very centre could not 
be considered due to severe saturation effects. In the central 0.75 pc, as in our study 
of Tr\,14, Brandner et al. (2008) could identify a similarly flat PDMF with $\Gamma\sim-0.5$.

However, it should be noted that we have to regard 
the effect of dynamical mass segregation in case of NGC\,3603\,YC and Westerlund\,1  
flattening the central PDMF because of the significantly higher density in these clusters. 
In contrast, no mass segregation was observed in the centre of Tr\,14 in the considered mass range. Based 
on a minimum spanning tree analysis, mass segregation was identified for stars more massive than 
10 M$_{\sun}$ \cite{sana}. They have analyzed the same MCAO data of Tr\,14 but including 
also the shallower MAD observations of the adjacent fields, such that the analyzed FoV 
encloses the central 2 arcmin. A comparable result has been reported for the 
ONC with mass segregation above 5 M$_{\sun}$ \cite {hillenbrand,allison}. 
Allison et al. (2009) pointed out that the observed high-mass segregation can be 
of dynamical origin which can happen on very short time-scales of less than 
1 Myr. The different degrees of segregation can be explained by the higher density 
of the ONC (about 15 times that of Tr\,14) reducing the time-scales on which mass segregation occurs.
In conclusion, a shallow PDMF as we observe in Tr\,14 is not an uncommon feature but 
observed in several massive young clusters, but its origin remains elusive.

\subsubsection{'Traditional' Mass Function}
\begin{figure}
\centering
\includegraphics[scale=0.5]{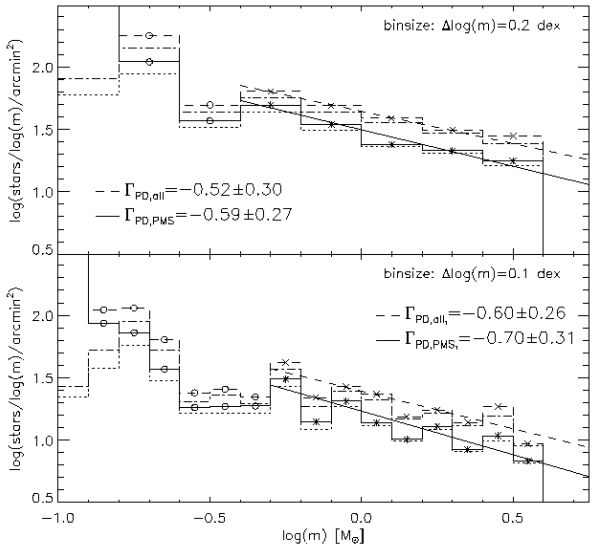}
\caption{MFs of Tr\,14 derived with the 'traditional' method. {\it Upper Panel:} 
The PDMF through star counts with bins of $\Delta log(m)=0.2$ dex. Above 0.5 M$_{\sun}$ 
we find ${\rm \Gamma_{PD,PMS}=-0.59\pm0.27}$ and ${\rm \Gamma_{PD,all}=-0.52\pm0.30}$. 
{\it Lower Panel:} The PDMF obtained via star counts with a bin size of 
$\Delta log(m)=0.1$ dex. Above 0.5 M$_{\sun}$ we derive similar slopes as for 
the $0.2$ dex-bins (${\rm \Gamma_{PD,PMS_1}=-0.70\pm0.31}$ and 
${\rm \Gamma_{PD,all_1}=-0.60\pm0.26}$).}
\label{kmf2}
\end{figure}

For comparison, we construct the PDMF also in the 'traditional' way, 
counting stars in logarithmic mass intervals with $\Delta log(m)=0.2$ dex and 
$\Delta log(m)=0.1$ dex. Stars are counted in each bin and normalised to the 
considered area. The results are corrected for incompleteness and the resulting 
PDMFs are shown in Fig.~\ref{kmf2}. Above 0.5 M$_{\sun}$, we 
fitted single-power law with the data points weighted by the individual Poisson 
errors. The corresponding slopes of are summarised in Table~\ref{slopes}. The 
derived power law slopes agree within the errors with the PDMF of Fig.~\ref{kmf}, 
although the PDMFs with 0.1 dex spacing appear steeper. 
Towards lower masses we also identify the drop of the mass function below 
0.5 M$_{\sun}$ followed by a steep increase below 0.25 M$_{\sun}$. Due to the 
sparsity of data points, we have not fitted the PDMF below 0.5 M$_{\sun}$.

\begin{table*}
\caption{Slopes of the different mass functions} \label{slopes}
\normalsize{
\begin{tabular}{ccccccc}
\hline
Mass range M$_{\sun}$ & Binning & ${\rm \Gamma_{PD,PMS}}$ & $\sigma_{{\rm \Gamma_{PD,PMS}}}$ & ${\rm \Gamma_{PD,all}}$ & $\sigma_{{\rm \Gamma_{PD,all}}}$
\\ \hline
$m_c-3.2$ & 1 star & -0.46 & 0.20 & -0.50 & 0.11\\
& 0.1 dex & -0.70 & 0.31 & -0.60 & 0.26 \\
& 0.2 dex & -0.59 & 0.27 & -0.52 & 0.30 \\
& & & & & \\
$0.25-m_c$ & 1 star & 0.73 & 0.59 & 0.63 & 0.32 \\
\hline
\hline
\end{tabular}
}
\end{table*}

\section{Summary}
We presented deep {\it H} and $K_S$ observations of the young massive cluster
Tr\,14, observed with MAD, the first MCAO system at the VLT.

{\it i) MAD (MCAO) in crowded regions:}\\We revealed 
an impressive performance of the AO system over a large FoV.  As part of the 
technical analysis we derived maps of Strehl ratios which show a low variation 
over the 68 arcsec$\times$68 arcsec FoV. Mean Strehl values are measured to 
6.0 and 5.9 per cent in {\it H}- and $K_S$-band, respectively. Simulating SCAO observations 
over the same FoV revealed smaller mean Strehls (1.6 per cent in {\it H} and 2.0 per cent in $K_S$). 
Although with lower maximum Strehl ratios (9.8 per cent in {\it H}-band, 12.6 per cent in $K_S$-band), 
when compared to SCAO systems, this shows a significant 
improvement of the spatial stability of the AO performance and demonstrates the 
opportunities offered by MCAO systems for wide-field AO corrected observations.

PSF photometry has been applied to derive properties of Tr\,14. The results show that 
the photometry is well estimated by using a second order variable PSF \cite[see also][]{sana}. 
Photometric calibration was accomplished using seeing-limited NTT--SofI observations 
\cite{ascenso07}. A spatially constant  ZP offset was found when compared to seeing-limited 
observations and used for photometric calibration. The resulting photometric catalogue 
comprises 972 (1347) sources down to very-low mass ($\lesssim0.1\ \rm{M}_{\sun}$) PMS stars. 
This doubles (triples) the number of detections when compared to seeing-limited observations 
of the same field.

{\it ii) $K_S$ vs. $H-K_S$ CMD of Tr\,14:}\\The photometric catalogue with almost 1,000 sources 
was used to create a $K_S$ vs. $H-K_S$ CMD of Tr\,14. It reveals a clear sequence of PMS 
stars, dozens of 'excess' sources and a sparsely populated MS. Comparing the CMD with PMS 
tracks by Siess et al. (2000), we find the CMD is best represented by an isochrone of 
$1\pm0.5$ Myr in age while hints for a somewhat older PMS population are observable. 
This places Tr\,14 at a distance of $2.3\pm0.4$ kpc with a foreground extinction of 
$A_{K_S}=0.38\pm0.03$ mag ($A_V=3.0\pm0.4$ mag). Hints for star formation over the last 
3 Myr support earlier claims of continuous star formation in the region and a 
slightly older halo population of the cluster \citep[e.g.][]{ascenso07}.

{\it iii) The LF and MF:}

The LF of Tr\,14 is derived based on the $K_S$ magnitudes. The observed 
turnover at $K_S=14.75$ mag is consistent with a cluster of the derived 
age of $\sim1$ Myr and the limiting mass at which Deuterium burning ceases.

We derived the PMS stellar PDMF that appears to be well represented by a broken 
power law between 0.25 and $3.2\ \rm{M}_{\sun}$. We found a PDMF that shows 
a change of the power law slope at $m_c=0.53\ \rm{M}_{\sun}$ comparable to that of 
a Kroupa--IMF. Such variations of the MF (flattening, turnover) of stellar systems 
are observed in other young clusters at similar stellar mass.
The corresponding power law slopes, however, show a shallower PDMF when compared to 
the Kroupa--IMF (Salpeter slope $\Gamma=-1.35$). In the intermediate- and low-mass 
regime we find slopes of $\Gamma_1=-0.50\pm0.11$ above $m_c$ and $\Gamma_2=0.63\pm0.32$ 
below $m_c$ down to $\sim0.25\ \rm{M}_{\sun}$. This reveals a deficiency in low-mass stars in 
the centre of Tr\,14. Furthermore, comparison with other young massive clusters shows 
similarly shallow PDMFs in the centres of young and massive clusters.

This study reveals the great improvement of AO corrections provided by MCAO systems 
in the case of wide field AO corrected observations. The stability of the correction for
atmospheric turbulence in combination with the correction over a few arcminutes 
provides an ideal combination for investigations that depend on high spatial resolution
instruments as well as a large FoV. Therefore, VLT--MAD gives a very promising
impression of the future prospects of observations with upcoming new telescopes and instruments.

\section*{Acknowledgments}
Our thanks go to the referee for the very constructive comments and discussion of our 
work leading to subsequent improvements of the paper. We would like to acknowledge 
the support by the Deutsches Zentrum f{\"u}r Luft- und Raumfahrt (DLR), 
F{\"o}rderkennzeichen 50 OR 0401 and the German Research Foundation (DFG) through 
the Emmy Noether grant STO 496/3-1.

\label{lastpage}


\begin{thebibliography}{}
\bibitem[Allard \& Freytag 2010]{allard} Allard F., Freytag B.\ 2010, Highlights of Astronomy, 15, 756 
\bibitem[Allison et al. 2009]{allison} Allison R.~J., Goodwin S.~P., Parker R.~J., de Grijs R., Portegies Zwart S.~F., Kouwenhoven M.~B.~N.\ 2009, \apjl, 700, L99 
\bibitem[Amorim et al. 2006]{amorim} Amorim A., et al.\ 2006, \procspie, 6269,
\bibitem[Ascenso et al. 2007]{ascenso07} Ascenso J., Alves J., Vicente S., Lago M.~T.~V.~T.\ 2007, \aap, 476, 199
\bibitem[Baraffe et al. 2002]{baraffe} Baraffe, I., Chabrier, G., Allard, F., \& Hauschildt, P.~H.\ 2002, \aap, 382, 563 
\bibitem[Bastian et al. 2010]{bastian} Bastian N., Covey K.~R., Meyer M.~R.\ 2010, \araa, 48, 339
\bibitem[Baumgardt \& Makino 2003]{baumgardt} Baumgardt H., Makino J.\ 2003, \mnras, 340, 227
\bibitem[Bertin \& Arnouts 1996]{bertin} Bertin E., Arnouts S.\ 1996, \aaps, 117, 393
\bibitem[Brandner et al. 2008]{brandner} Brandner W., Clark J.~S., Stolte A., Waters R., Negueruela I., Goodwin S.~P.\ 2008, \aap, 478, 137 
\bibitem[Campbell et al. 2010]{campbell} Campbell M.~A., Evans C.~J., Mackey A.~D., Gieles M., Alves J., Ascenso J., Bastian N., Longmore A.~J.\ 2010, \mnras, 405, 421
\bibitem[Cardelli et al. 1989]{cardelli} Cardelli J.~A., Clayton G.~C., Mathis J.~S.\ 1989, \apj, 345, 245 
\bibitem[Carraro et al. 2004]{carraro} Carraro G., Romaniello M., Ventura P., Patat F.\ 2004, \aap, 418, 525
\bibitem[Cresci et al. 2005]{cresci05} Cresci G., Davies R.~I., Baker A.~J., Lehnert M.~D.\ 2005, \aap, 438, 757
\bibitem[Da Rio et al. 2009]{dario09} Da Rio, N. Gouliermis D.~A., Henning T., 2009, \apj, 696, 528
\bibitem[Devillard 2001]{devillard} Devillard N.\ 2001, Astronomical Data Analysis Software and Systems X, 238, 525 
\bibitem[Gennaro et al. 2011]{gennaro} Gennaro M., Brandner W., Stolte A., Henning T.\ 2011, \mnras, 162 
\bibitem[Harayama et al. 2008]{harayama} Harayama Y., Eisenhauer F., Martins F.\ 2008, \apj, 675, 1319 
\bibitem[Hillenbrand 1997]{hillenbrand} Hillenbrand L.~A.\ 1997, \aj, 113, 1733 
\bibitem[Kroupa 2001]{kroupa} Kroupa P.\ 2001, \mnras, 322, 231
\bibitem[Lada \& Lada 2003]{ladalada} Lada C.~J., Lada E.~A.\ 2003, \araa, 41, 57
\bibitem[Ma{\'{\i}}z Apell{\'a}niz \& {\'U}beda 2005]{maiz} Ma{\'{\i}}z Apell{\'a}niz J., \& {\'U}beda L.\ 2005, \apj, 629, 873
\bibitem[Ma{\'{\i}}z Apell{\'a}niz 2009]{maiz09} Ma{\'{\i}}z Apell{\'a}niz J.\ 2009, \apss, 324, 95
\bibitem[Marchetti et al. 2003]{marchetti03} Marchetti E., Ragazzoni R., Diolaiti E.\ 2003, \procspie, 4839, 566
\bibitem[Marchetti et al. 2004]{marchetti04} Marchetti E., et al.\ 2004, \procspie, 5490, 236
\bibitem[Marchetti et al. 2007]{marchetti07} Marchetti E., et al.\ 2007, The Messenger, 129, 8
\bibitem[Marigo et al. 2008]{marigo} Marigo P., Girardi L., Bressan A., Groenewegen M.~A.~T., Silva L., Granato G.~L.\ 2008, \aap, 482, 883
\bibitem[Massey \& Johnson 1993]{massey93} Massey P., Johnson J.\ 1993, \aj, 105, 980
\bibitem[Momany et al. 2008]{momany} Momany Y., Ortolani S., Bonatto C., Bica E., Barbuy B.\ 2008, \mnras, 391, 1650 
\bibitem[Naylor 2009]{naylor} Naylor T.\ 2009, \mnras, 399, 432 
\bibitem[Nelan et al. 2004]{nelan04} Nelan E.~P., Walborn N.~R., Wallace D.~J., Moffat A.~F.~J., Makidon R.~B., Gies D.~R., Panagia N.\ 2004, \aj, 128, 323 
\bibitem[Rathborne et al. 2002]{rathborne02} Rathborne J.~M., Burton M.~G., Brooks K.~J., Cohen M., Ashley M.~C.~B., Storey J.~W.~V.\ 2002, \mnras, 331, 85
\bibitem[Rochau et al. 2010]{rochau} Rochau B., Brandner W., Stolte A., Gennaro M., Gouliermis D., Da Rio N., Dzyurkevich N., Henning T.\ 2010, \apjl, 716, L90 
\bibitem[Rousset et al. 1990]{rousset90} Rousset G., Fontanella J.~C., Kern P., Gigan P., Rigaut F.\ 1990, \aap, 230, L29
\bibitem[Sagar \& Richtler 1991]{sagar} Sagar R., Richtler T.\ 1991, \aap, 250, 324 
\bibitem[Salpeter 1955]{salpeter} Salpeter E.~E.\ 1955, \apj, 121, 161
\bibitem[Sana et al. 2010]{sana} Sana H., Momany Y., Gieles M., Carraro G., Beletsky Y., Ivanov V.~D., de Silva G., James G.\ 2010, \aap, 515, A26 
\bibitem[Siess et al. 2000]{siess00} Siess L., Dufour E., Forestini M., 2000, \aap, 358, 593
\bibitem[Smith 2006]{smith06} Smith N.\ 2006, \mnras, 367, 763
\bibitem[Stetson 1990]{stetson} Stetson P.~B., 1990, \pasp, 102, 932
\bibitem[Stolte et al. 2002]{stolte} Stolte A., Grebel E.~K., Brandner W., Figer D.~F.\ 2002, \aap, 394, 459 
\bibitem[Stolte et al. 2006]{stolte06} Stolte A., Brandner W., Brandl B., Zinnecker H.\ 2006, \aj, 132, 253 
\bibitem[Sung \& Bessell 2004]{sung} Sung H., Bessell M.~S.\ 2004, \aj, 127, 1014
\bibitem[Tapia et al. 2003]{tapia03} Tapia M., Roth M., V{\'a}zquez R.~A., Feinstein A.\ 2003, \mnras, 339, 44
\bibitem[Vazquez et al. 1996]{vazquez96} Vazquez R.~A., Baume G., Feinstein A., Prado P.\ 1996, \aaps, 116, 75
\bibitem[Walborn et al. 2002]{walborn} Walborn, N.~R., et al. 2002, \aj, 123, 2754
\bibitem[Zinnecker \& McCaughrean 1991]{zinn+mcc} Zinnecker H., McCaughrean M.\ 1991, \memsai, 62, 761 
\end{thebibliography}
\end{document}